\documentclass{aastex631}

\shorttitle{Sudden outburst and inner disk wall destruction}
\shortauthors{Xie et al.}
\usepackage{multirow}
\graphicspath{{./}{figures/}}

\newcommand{\uat}[2]{\href{http://vocabs.ands.org.au/repository/api/lda/aas/the-unified-astronomy-thesaurus/current/resource.html?uri=http://astrothesaurus.org/uat/#1}{#2  (#1)}}
\begin{document}

\title{JWST captures a sudden stellar outburst and inner disk wall destruction}

\newcommand{\affilLPL}{\affiliation{Lunar and Planetary Laboratory, The University of Arizona, Tucson, AZ 85721, USA; \url{cyxie@arizona.edu}}}

\correspondingauthor{Chengyan Xie}
\email{cyxie@arizona.edu}

\author[0000-0001-8184-5547]{Chengyan Xie}
\affilLPL

\author[0000-0001-7962-1683]{Ilaria Pascucci}
\affilLPL

\author{Dingshan Deng}
\affilLPL

\author{Naman S. Bajaj}
\affilLPL

\author{Richard Alexander}
\affiliation{School of Physics and Astronomy, University of Leicester, University Road, Leicester LEI 7RH, UK}

\author{Andrew Sellek}
\affiliation{Institute of Astronomy, Madingley Road, Cambridge CB3 0HA, UK}
\affiliation{Leiden Observatory, Leiden University, 2300 RA Leiden, The Netherlands}

\author{\'Agnes K\'osp\'al}
\affiliation{Konkoly Observatory, HUN-REN Research Centre for Astronomy and Earth Sciences, MTA Centre of Excellence, Konkoly-Thege Mikl\'os \'ut 15-17, 1121 Budapest, Hungary}
\affiliation{Institute of Physics and Astronomy, ELTE E\"otv\"os Lor\'and University, P\'azm\'any P\'eter s\'et\'any 1/A, 1117 Budapest, Hungary}
\affiliation{Max Planck Institute for Astronomy, K\"onigstuhl 17, 69117 Heidelberg, Germany}

\author{Giulia Ballabio}
\affiliation{Astrophysics Group, Imperial College London, Blackett Laboratory, Prince Consort Road, London SW7 2AZ, UK}

\author{Uma Gorti}
\affiliation{NASA Ames Research Center, Moffett Field, CA 94035, USA}
\affiliation{Carl Sagan Center, SETI Institute, Mountain View, CA 94043, USA}

%\author{Andras Gaspar}
%\affiliation{Steward Observatory, The University of Arizona, Tucson, AZ 85721, USA}

%\author{Jane Morrison}
%\affiliation{Steward Observatory, The University of Arizona, Tucson, AZ 85721, USA}

\begin{abstract}
We analyze JWST/MIRI observations of T~Cha, a highly variable ($\Delta V \sim$3-5\,mag) accreting Sun-like star surrounded by a disk with a large ($\sim 15$\,au) dust gap. We find that the JWST mid-infrared spectrum is significantly different from the  {\it Spitzer} spectrum obtained 17 years before $-$ the emission at short wavelengths ($5-10 \mu m$) has decreased by $\sim 2/3$ while that at longer wavelengths ($15-25 \mu m$) has increased by up to a factor of $\sim 3$. The JWST spectrum is contemporary with a fairly constant higher optical emission captured by the All Sky Automated Survey. By analyzing and modelling both SEDs, we propose that JWST caught the star during an outburst that partly destroyed and significantly reduced the height of  the asymmetric inner disk wall responsible for the high optical variability and lower $15-25$\,\micron\ emission during the {\it Spitzer} time. The dust mass lost during this outburst is estimated to be comparable ($\sim 1/5$) to the upper limit of the total micron-sized dust mass in the inner disk of T~Cha now. Monitoring this system during possible future outbursts and more observations of its quiescent state will reveal if the inner disk can be replenished or will continue to be depleted and vanish.

\end{abstract}

\keywords{ \uat{235}{Circumstellar disks}, \uat{1300}{Protoplanetary disks}, \uat{1257}{Planetary system formation}, \uat{786}{Infrared astronomy}, \uat{1290}{Pre-main sequence stars} }
%%%%%% Here is the structure of the paper
%% Introduction to discuss the properties of TCha from literature
%% Section 2. TCha parameters; the meaning of seesaw (what we want to achieve in our fitting: Decrease in inner disk emission and increase in outer disk emission);
%% Section 3, Models and analysis
%% In section 3, we should include: other possible explanations; fixed parameters selection; the two typical results from symmetric and non-symmetric model
%% Section 4, Discussion.
%% In section 4, we should include: hints of inner disk evolution. 
%%%%%

\section{Introduction} \label{sec:intro}
Gas-rich dust disks around young stars are the birthplaces of planets, yet many questions on how planets form remain unanswered, highlighting the critical need for further studies of their birth environments.
These planet-forming disks were first identified via infrared excess emission on top of the stellar photosphere. 
Among them, those having low near-infrared (NIR) excess, but pronounced mid- and far-infrared (IR) excess are particularly interesting as this type of spectral energy distribution (SED) indicates a discernible diminishment in dust mass in the inner disk. Such disks are called transitional disks and thought to be  the transition stage from optically thick full disks to dissipated Class III sources \citep[see][for a review]{Espaillat14}{}{}. 
While star-driven photoevaporation \citep[e.g.,][]{Alexander14}{}{} might explain the inside-out clearing of a  subset of these objects, it is clear that those transition disks with large mass accretion rates and/or large dust cavities require a different process \citep[see Fig.~6 in][]{Ercolano17}{}{}. In a few cases planet formation is known to be the culprit of the dust clearing \citep[e.g.,][]{Keppler18,Christiaens24}{}{} 
%Several mechanisms can cause this inner dust clearing, including accretion, disk winds, and the presence of planets. Viscous accretion arise from magneto-rotational instability driven turbulence which drives a inward flow of gas and small dust grains (micron size) towards the central star while transporting angular momentum outwards \citep[e.g.,][]{Espaillat14,Hartmann16}{}{}. Photoevaporation models show radiation from the central star heats the disk surface, establishing a thermal wind. As the accretion rate decreases over time, at a certain point falls below the wind mass loss rate. Photoevaporation will take over further evolution of the disk, quickly erode the disk from inside out \citep[see][for a review]{Ercolano17}{}{}. In addition, planets present or forming in a disk can carve out a gap and form a local pressure maxima just outside the gap, trapping grains and result in ring structures \citep[e.g.,][]{Rafikov02,Bae23}{}{}. 

The central stars of the disks (also called T Tauri Stars) are young, accreting and variable at optical wavelengths. The Spitzer Space Telescope (hereafter {\it Spitzer}) revealed variability also at mid-IR wavelengths from their dust disks. An interesting subset of transition disks shows ``seesaw'' variability at mid-IR band, where
the emission at shorter wavelengths varies inversely with the emission at longer wavelengths \citep[e.g.,][]{Muzerolle09,Espaillat11,Flaherty12,Kospal12,Espaillat24}{}{}. It was proposed that changes of the height of the optically thick inner edge or wall located close to the dust destruction radius (where the dust is hot and emits primarily at NIR wavelengths) could lead to the changes in shadowing of the outer disk (which emits at longer wavelengths), thus resulting in the seesaw effect \citep[see Fig.~7 of ][for an illustration]{Espaillat14}{}{}. The inner dust disk wall height variation can be caused by many factors including the dynamical interaction between the stellar magnetic field and the disk \citep[e.g.,][]{Bouvier99,Lai08}{}{}, perturbations in disks by planets, or turbulence \citep[e.g.,][]{Espaillat14}{}{}. Besides {\bf changes in the} inner disk wall height, a warp in the disk can also result in seesaw changes in the SED \citep[e.g.,][]{Nixon10}. 

%Jame Webb Space Telescope (JWST), launched in December 2021, is a high-resolution and high-sensitivity space telescope designed to conduct IR study. The mid-infrared instrument (MIRI) Medium-resolution spectrometer (MRS) on JWST not only reveals a lot more weak features on the spectra with higher resolution and sensitivity, but also set up a much longer timeline ($\sim 15$ years) when compared with previous Spitzer observations, enabling the  analysis of the long-term time variability of disks.
The launch of the James Webb Space Telescope (JWST) in December 2021 opens new opportunities for investigating variability at mid-IR wavelengths again after {\it Spitzer}, including more long-term ($\sim$ 15 years) variability of disks by combining observations from both telescopes.
%The launch of the James Webb Space Telescope (JWST) in December 2021 set up a much longer timeline ($\sim 15$ years) when compared with previous {\it Spitzer} observations, enabling the analysis of the long-term time variability of disks.
In this paper, we present an analysis of the JWST MIRI/MRS spectrum of a transition disk system around T Chamaeleontis \citep[hereafter T~Cha, see][for details about the observations]{Bajaj24}. T~Cha is a G8V T Tauri star \citep{Alcala93} located at a distance of $\sim$ 102.7\,pc \citep{Gaia3} in the $\epsilon$-Cha star forming region \citep{Murphy13}.  We compare the 2022 MIRI spectrum with a previous {\it Spitzer} IRS spectrum obtained in 2005 and other optical to millimeter photometric data. 
We show that the SED of T~Cha has significantly changed compared to the {\it Spitzer} epoch, $\sim$17 years earlier. By modelling this change along with ASAS-SN monitoring, we found that T~Cha is likely to have partly lost its inner dust wall in 2021 and regained it in 2023. The SED of T~Cha is discussed in section~\ref{sec:obs}. The model to fit the observations is presented in section~\ref{sec:model}, while the physical meaning and the implications of our model are outlined in section~\ref{sec:dis}. We summarize our findings in section~\ref{sec:sum}.

\section{T Cha and its variable spectrum} \label{sec:obs}
T~Cha has been long known to be surrounded by a circumstellar disk \citep[e.g.,][]{Alcala93}{}{}. \cite{Brown07} observed T~Cha with  {\it Spitzer} and inferred a large dust gap ($\sim$15~au) from modeling its SED. The gap was confirmed by follow-up NIR interferometry \citep{Olofsson11,Olofsson13} and high-resolution millimeter imagery \citep{Hendler18}. The latter observations reveal the existence of an unresolved inner disk with radius $<$1\,au, and
an outer disk beyond $\sim$20~au with an inclination of $\sim$70$^\circ$. T~Cha is also known to be accreting disk gas with an accretion rate ($\dot{M}$) varying from $\sim 1 \times 10^{-9} $ to $ 3\times 10^{-8} M_{\odot}\,{\rm yr}^{-1}$ \citep{Schisano09,Cahill19}. Continued accretion and $^{12}$CO gas emission maps obtained with ALMA \citep{Wolfer23} suggest that there is no significant gap in the gas.
%The range of $\dot{M}$ suggests} that there is no significant gap in the gas, which is further supported by  a $^{12}$CO gas emission map \textbf{obtained with ALMA} \citep{Wolfer23}. 
Stellar parameters of T~Cha that are relevant for this study are summarized in Table~\ref{tab:para}.
\begin{deluxetable*}{c|cc}
%\tablenum{1}
\tablecaption{T~Cha stellar properties relevant to our study\label{tab:para}}
\tablewidth{0.99\textwidth}
\tablehead{
%\multicolumn{3}{c}{Fixed parameters}\\
%\hline
Property & Value & Ref. \\
}
\startdata
D (pc) & $102.7 \pm 0.4 $ & 1\\
SpT & G8 & 2 \\
log$_{10}$($L_*$) ($L_\odot$) &$0.13$ & 3\\
$M_*$ ($M_\odot$) & $1.5$& 2\\
$R_*$ ($R_\odot$) & $1.3$& 3\\
log$_{10}$($\dot{M})$($M_{\odot}$/yr)& \textbf{-7.5$\sim$-9.0} &4, 5\\
T$_{*}$(K)& 5600 & 6\\
\enddata

\tablerefs{1.~\cite{Gaia3}; 2.~\cite{Alcala93}; 3.~\cite{Olofsson11}; 4.~\cite{Schisano09} 5. ~\cite{Cahill19} 6.~\cite{Brown07}}
\end{deluxetable*}

\subsection{Variability of T~Cha} \label{sec:obs:oldobs}

\begin{figure*}[htb!]
    \centering
	\includegraphics[width=0.99\textwidth]{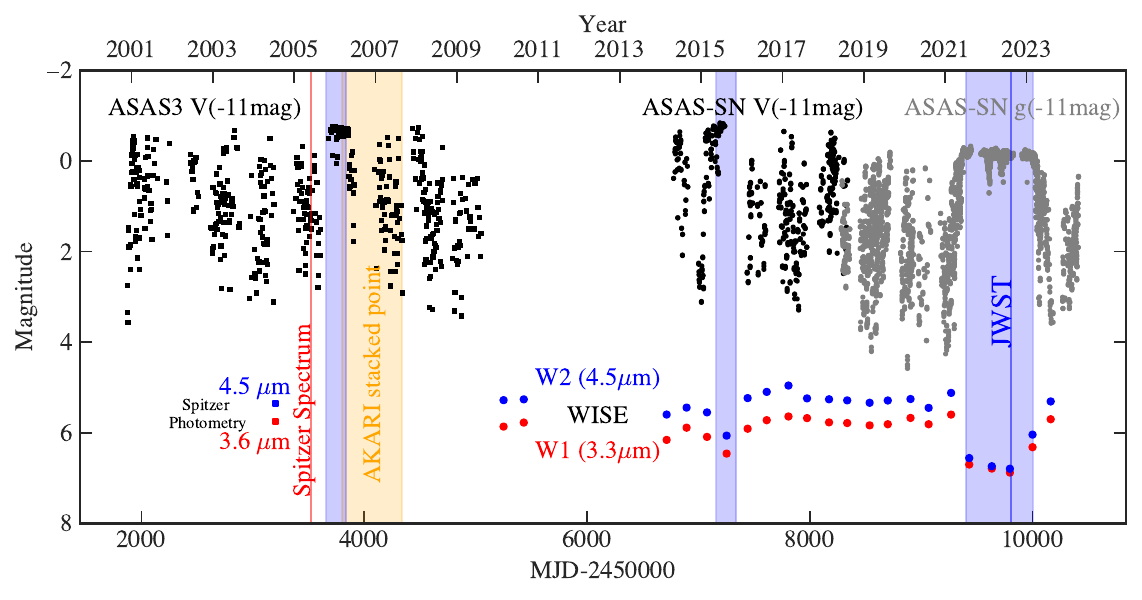}
    \caption{Optical and mid-IR light curves of T~Cha. The lines and shaded areas indicate the date when some of the photometric points we analyzed in this paper were taken. Blue shaded areas indicate the period when T~Cha has less variable but brighter optical emission. %\sout{ The `LCOGT and Andicam' is from \cite{Walter18}} \ilaria{Remove the green band}.
    %: \cite{Alcala93} (black dots with dased lines); 
    %\ilaria{include detailed description of the SED. You should also mention what are the strong emission lines, ArII and NeII and that they will be discussed in Bajaj et al.}
    }
    \label{fig:Photometry}
\end{figure*}

Like many other young stellar objects, T~Cha shows short-term photometric variability at optical and NIR wavelengths \citep[e.g.,][]{Covino92,Alcala93,Walter18}. %However, its optical variability is much larger than the average T Tauri variability \citep[e.g.,][and Fig~\ref{fig:Photometry}]{Osterloh96,Rigon17}{}{}
According to previous studies, the optical variability of T~Cha amounts to a change in visual magnitude of $\sim 3$ \citep[V magnitude from 10 to 13, e.g.,][]{Alcala93,Walter18}{}{}. In this paper, we also incorporate optical monitoring from the All Sky Automated Survey (including ASAS3 during 2000-2009 and ASAS-SN from 2014), and find that at times T~Cha can be as dim as $V \sim 15$\,mag. This means that the total change in visual magnitude can be up to $\sim 5$ (see Fig~\ref{fig:Photometry}). Such optical variability is much larger than the average T Tauri variability \citep[e.g.,][]{Osterloh96,Rigon17}{}{}.
%\footnote{In previous papers the variability is $\sim 3$ \citep[V magnitude from 10 to 13, e.g.,][]{Alcala93,Walter18}{}{}, but the ASAS3 and ASAS-SN data shows in some cases T~Cha can be as dim as $V \sim 15$ magnitude (see Fig~\ref{fig:Photometry})}. 
About half of the optical variability shows a periodic behavior in $\sim$ 3.3 days \citep[e.g.,][]{Mauder75,Walter18}{}{}, while there is also significant random scatter superimposed on the periodic variability. The emission from all bands (from B to K band) increases or decreases together, with redder colors when the emission fades \citep[e.g.,][and Figure 2]{Walter18}. 
%\sout{During {\bf these studied stages} (when $\Delta V \sim 3$ mag, e.g., the green shaded region \textbf{indicated with ``LCOGT and Andicam''} in Fig.~\ref{fig:Photometry}), around half of the variability ($\Delta V \sim 1.5$ mag) shows a periodic behavior in $\sim$ 3.3 days \citep[e.g.,][]{Mauder75,Walter18}{}{}, while there is also significant random scatter superimposed on the periodic variability \citep[see Figure~5 of][]{Walter18}{}{}.} \ilaria{COMMENT: THIS IS VERY CONFUSING. IF YOU LOOK AT FIGURE 2 YOU SEE THE SAME BEHAVIOUR DURING THE NON VARIABLE STAGE AND LET'S REMOVE THE GREEN ADCICAM STRIPE FROM FIGURE 1, SEE MY CHANGES ABOVE.} 
%In addition to these, during some special periods T~Cha also shows relatively weak optical variability, accompanied with consistent higher optical emission and less mid-IR emission (see the blue shaded regions in Fig.~\ref{fig:Photometry}). 
%The photometric points also show a large random scatter despite the short-term periodicity. %\sout{ Optical variability in young stellar objects can be caused by many mechanisms\citep[e.g.,][]{Bouvier99}{}{}, including stellar spots and photosphere obscuration. }

Optical and NIR variability of young stellar objects can arise from many photospheric processes including outbursts, magnetic cool spots and unsteady accretion hot spots \citep[e.g.,][]{Herbst94,Fernandez96,Fischer23}{}{}. Outbursts are typically one-time events and are not likely to be the origin of the short-term ($\sim 3$ days) variability of T~Cha \citep[e.g.,][]{Lorenzetti12,Findeisen13,Fischer23}{}{}. Hot spots are typically caused by the accretion onto the star. They can lead to brighter emissions above the stellar photospheric model and are generally irregular on timescales as short as hours \citep[e.g.,][]{Herbst94}{}{}, which can explain those photometric points that are randomly scattered.
%the random scattering of optical photometric points for T~Cha. 
However, periodic variations in the stellar photosphere like cool spots are typically weak (largest $\Delta V \sim0.8$ mag), and are unable to explain the periodic optical variability seen in most epochs of T~Cha ($\Delta V \gtrsim1.5$ mag). 
The observation that the star becomes redder as it fades, along with the  short $\sim 3.3$ days periodicity,  points to obscuration from a very close inner disk located at $\sim$0.05~au. This location is within the range of possible magnetic truncation radii \citep[$\sim5-10~R_*$ or $\sim0.03-0.06$~au for T~Cha e.g.,][]{Gullbring98,Bovier07a}. This scenario is consistent with the brightening of  optical emission lines like H$\alpha$ and [O\,{\scriptsize I}] when the star fades due to contrast enhancement relative to the photosphere \citep{Schisano09}.

Contrary to the strong optical and NIR variability, the photometric points at longer than 3~$\mu m$ obtained from 1970s to 2010, including mid-IR \citep[$\sim$ 3-30$\mu$m, e.g.,][]{IRAS94,MSX03,Spitzer03,Akari10,WISE12}{}{} and FIR \citep[$\sim$ 50-500 $\mu$m, e.g.,][]{IRAS94,Akari10,Cieza11}{}{}, show much less variability \citep[by a factor of $<5\%$ for 3 $-$ 10$\mu m$ and $<50\%$ for 10-500$\mu m$ in flux, e.g.,][]{Schisano09,Cieza11}. This indicates the emission from the outer disk, which is cooler and emits longer wavelength photons compared to the star, was mostly stable during previous decades. 

\begin{figure*}[htb!]
    \centering
	\includegraphics[width=0.99\textwidth]{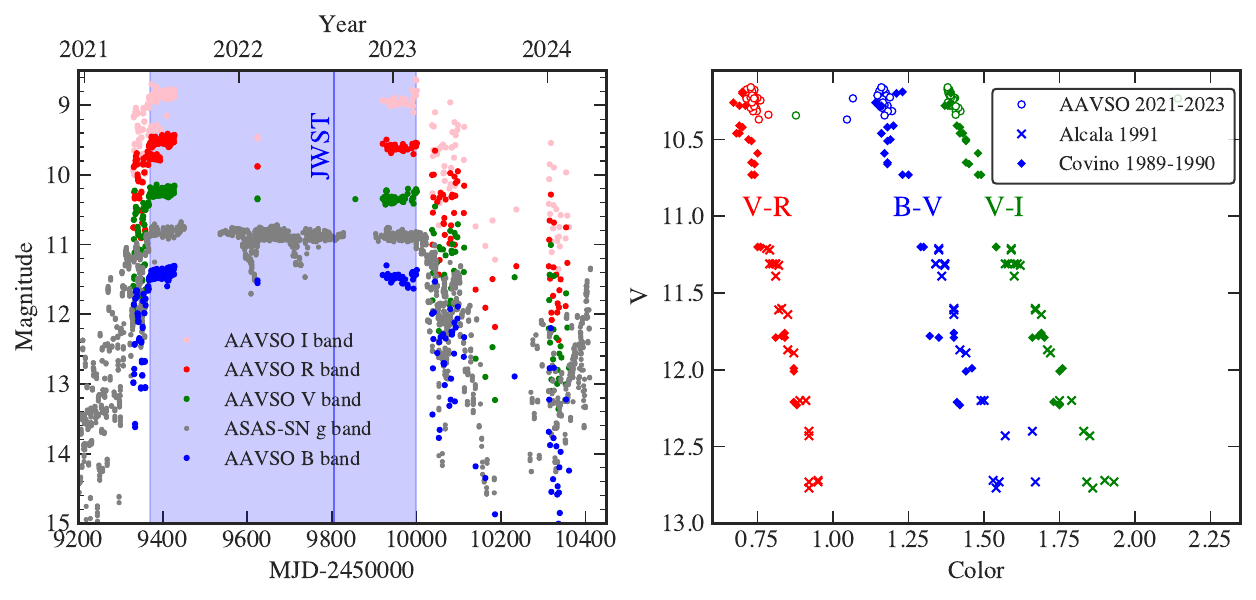}
    \caption{\textbf{Left Panel:} AAVSO photometric data (colored) plotted with the ASAS-SN data (gray). The JWST observation time and blue shaded area are plotted the same as Fig~\ref{fig:Photometry}. The emissions of T~Cha in all optical and NIR bands are less variable and brighter during the same period between 2021 and 2023. 
    \textbf{Right Panel:} Color magnitude diagram from AAVSO data in the blue shaded period (circles) compared with the photometry during 1991 \citep{Alcala93} (crosses) and 1989-1990 \citep{Covino92} (dimonds). When T~Cha gets brighter, it gets bluer. 
    }
    \label{fig:AAVSO}
\end{figure*}

Recent WISE measurements, however, reveal a drop in flux of T~Cha at 3 and 5 $\mu$m in 2015 and 2021 (by a factor of $\sim 40\%$ and $\sim 70\%$ in flux, respectively). Both drops are accompanied by a brighter and less variable optical emission\footnote{Similar optical behavior was also seen in 1989 \citep{Covino92,Alcala93} and 2006 (blue shaded area in Fig.~\ref{fig:Photometry}), but no contemporary IR observation was taken.} seen in ASAS surveys (Fig.~\ref{fig:Photometry}).
The larger change in 2021 also persists for longer, with both the decrease in WISE bands and the constant brighter optical emission lasting for $\sim 2$ years. Observations from The American Association of Variable Star Observers (AAVSO) shown in Fig~\ref{fig:AAVSO} identified the same optical and NIR behavior in different bands and comparison with historical photometry \citep[e.g.,][]{Covino92,Alcala93} positions T~Cha during this time on the brightest optical points and smallest color indices. 
%\sout{Comparing the \sout{emission during the} 2021 \ilaria{photometry} \sout{optically brighter period} with historical photometry \citep[e.g.,][]{Covino92,Alcala93} \ilaria{in Fig.2}  shows that T~Cha gets bluer when it gets brighter, implying less extinction and possibly more accretion during these periods.}
The change in both optical and IR bands indicates a possible change in both the inner disk (e.g., less extinction) and the stellar photosphere (e.g., more accretion  or an outburst).
%During these periods, \textbf{there is more emission on the blue (3.3$\mu m$) than on the red (4.5$\mu m$) at IR wavelengths}, suggesting \textbf{extinction is} lower. 

%In addition to these, during some special periods T~Cha also shows relatively weak optical variability, accompanied with consistent higher optical emission and less mid-IR emission (see the blue shaded regions in Fig.~\ref{fig:Photometry}). 

%\cyxie{Maybe put a subsection here?}
%\subsection{A change in the mid-IR SED of T~Cha}
\subsection{JWST MIRI/MRS reveals mid-IR changes for T~Cha} \label{sec:obs:newobs}
T~Cha was observed with the Medium Resolution spectrometer \citep[MRS,][]{Wells15} of the JWST Mid-Infrared Instrument \citep[MIRI,][]{Rieke15}  as part of the Cycle~1 General observer program (ID: 2260, PI: I. Pascucci). It was observed on Aug 13-14, 2022 for 3.24 hrs on-source and 4.48 hrs including the overheads. In short, the JWST Calibration Pipeline \citep{Bushouse23} version 1.11.2 and Calibration Reference Data System (CRDS) context \texttt{jwst\_1100.pmap} was used to process the data up to stage 2 resulting in datacubes for every exposure, channel, and sub-band. We minimize the fringes by using a fringe flat field and applying the residual\_fringe step (Kavanagh et al. in prep.), which are included in the JWST calibration pipeline package as part of \texttt{Spec2Pipeline}. 
Further details on the data reduction can be found in \cite{Bajaj24}. 

\begin{figure*}[htb!]
    \centering
	\includegraphics[width=0.59\textwidth]{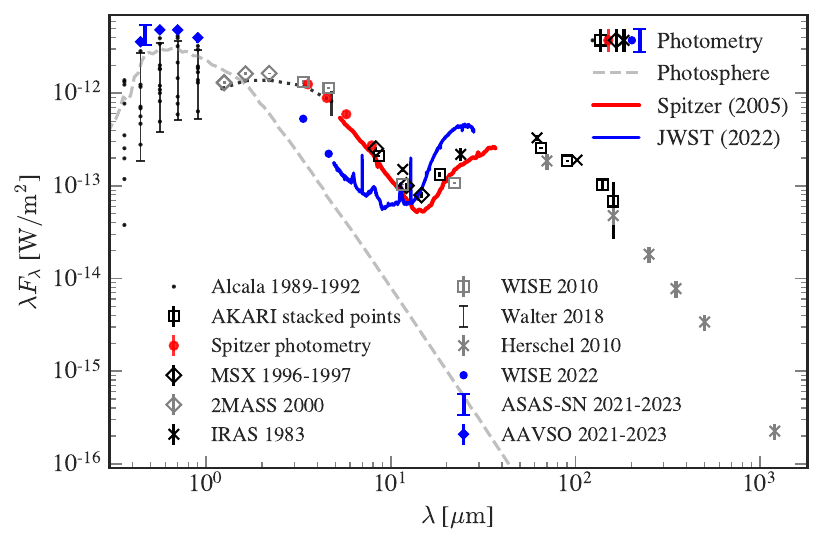}
    \caption{Overview of the SED of T~Cha. The grey dased line is 5600~K stellar photosphere emission scaled to the distance to T~Cha. The red and blue solid lines are {\it Spitzer}/IRS SED and JWST MIRI/MRS spectra, respectively. The two strong lines in the JWST spectrum are [ArII] and [NeII], and the broad features are PAH features. 
    The photometric points in optical bands are from \cite{Alcala93,Walter18} and AAVSO, and the IR photometric points are from various papers (see Table~\ref{app:tab:ref}). We highlight some points in blue to indicate they are taken contemporary with the JWST observation (see the blue shaded region next to the JWST line in Fig~\ref{fig:Photometry}). The near and mid-IR points connected with dotted lines indicate one contemporary observation at JHKLM bands \citep{Alcala93}.
    %\cyxie{I feel the points are from too many different surveys, do we want to indicate every survey as \cite{Schisano09}, or just say it briefly?}
    %: \cite{Alcala93} (black dots with dased lines); 
    %\ilaria{include detailed description of the SED. You should also mention what are the strong emission lines, ArII and NeII and that they will be discussed in Bajaj et al.}
    }
    \label{fig:Overview}
\end{figure*}

Fig~\ref{fig:Overview} shows the SED of T~Cha. All the plotted photometric points and spectra are dereddened with $A_V=1.3$\,mag \citet[][estimated from the photometry near the maximum brightness]{Schisano09} and $R_V=3.1$. This accounts for interstellar extinction while additional extinction from the disk will be properly incorporated in our modeling (see Sections~\ref{sec:model}). Table~\ref{app:tab:ref} provides the references for the data plotted in Fig.\ref{fig:Overview} and a downloadable table with the photometric points. The photospheric model for a G8 star from BT-Settl model is scaled to the distance of T~Cha. 
Note that the MIRI-MRS spectrum (in blue) is significantly different from the previous {\it Spitzer}/IRS spectrum \citep[in red,][]{Brown07} and most other mid-IR photometric points \citep[e.g.][see fig~\ref{fig:Overview}]{MSX03,Spitzer03,WISE12}, but is consistent with the recent (2021-2023) WISE photometric points (blue dots). While the shorter portion ($<$10$\mu$m) of the SED decreased in flux,  the longer portion ($>$15$\mu$m) increased, both by a factor of up to $\sim 3$. This large change at mid-IR wavelengths implies the disk around T~Cha has experienced a significant change in 2021, more specifically this appears as a seesaw behavior with a pivot point at $\sim$12 $\mu$m (see section~\ref{sec:intro}). 
%Another significant phenomenon in this period is the constant optical brightening, up to $\sim$2 times higher than the 5600~K photospheric model(see the blue line at $\sim 0.48 \mu m$ in Fig~\ref{fig:Overview}).}
Among the mid-IR variable sources studied with {\it Spitzer}, LRLL~31 is the most similar  to T~Cha. LRLL31 not only shares a similar spectral type to TCha but also has a comparable mass accretion rate, disk inclination, and variation in mid-IR flux \citep{Muzerolle09}.  The seesaw behavior of LRLL~31 was  explained as a change in the inner disk wall \citep{Flaherty10,Bryan19}.
%\sout{\cyxie{A similar behavior is seen in another mid-IR variable source, LRLL~31, studied with {\it Spitzer}. Compared with T~Cha, LRLL~31 not only has similar spectral type, mass accretion rate and disk inclination, but the magnitude of change in flux at mid-IR wavelengths is comparable as well \citep{Muzerolle09}.  The seesaw behavior of LRLL~31 was  explained as a change in the inner disk wall \citep{Flaherty10,Bryan19}}}
%Among the mid-IR variable sources \textbf{studied with {\it Spitzer}}, LRLL~31 is the most similar one to T~Cha. They not only have similar spectral type, mass accretion rate and disk inclination, but the magnitude of change in flux at mid-IR wavelengths is comparable as well \citep{Muzerolle09}.  The seesaw behavior of LRLL~31 was explained as a change in the inner disk wall \citep{Flaherty10,Bryan19}. 
We also note that there are stronger ionic lines, which trace a disk wind, and PAH features, which are excited by UV radiation \citep[e.g.,][]{Geers06}, in the JWST spectrum of T~Cha compared to the {\it Spitzer} spectrum. This hints at a stronger UV radiation and a higher disk mass loss rate, which can be associated with an increase in accretion activity during this period \cite[as discussed in][]{Bajaj24}.

%Though resemble in trends, T~Cha is special in several ways: 1. Compared to the typical $<40\%$ changes in flux in \cite{Espaillat11}, T~Cha shows a much stronger change in flux; 2. The flux change timescale of the sources studied by \cite{Espaillat11} are typically in weeks, while for T~Cha it's $\sim 15$ years.
%The timescale of the change in T~Cha is relatively longer than the  \ilaria{These 3 statements need to be strenghtened: in 1. what is the typical change in flux in Espaillat? You need to provide the value so taht the reader can compare with the factor of 2 for TCha. In 2. what is the timescale of variability for the Espaillat sources? Why is 3.? Because the timescale is longer? I would drop 3. NOTE: that there is  asource very similar to TCha (see Muzerolle+2009 and Bryan+2019), this source should be mentioned too}

%The magnitude of change in flux of LRLL~31 is large and comparable to T~Cha, and this see-saw behavior is explained by the change of inner disk wall. 

\section{Models and Analyses} \label{sec:model}

In this section, we discuss disk models that can reproduce the overall SED of T~Cha as well as the change between the {\it Spitzer}/IRS and the JWST MIRI-MRS spectra. Specifically, these models need to account for: 1. The mid-IR seesaw behavior, which indicates less emission from the hot inner disk and more emission from the cool outer disk, from the {\it Spitzer} to the JWST spectrum.
2. The periodic part of the optical variability which is too strong to be only due to stellar variability, and is significantly reduced in 2021.

\subsection{Model setup}
%\ilaria{This section needs to be organized better, it is very difficult to read. I have added some subsections in italics, relevant text should be grouped in tehse subsections}

We use the DiskMINT wrapper \citep{Deng23} to model the SED of T~Cha with the radiative transfer code RADMC-3D \citep{RADMC3D12}. 
In our model, we adopt the stellar radius and mass $R_\star=1.3R_\odot$, $M_*=1.5M_\odot$ from \cite{Alcala93,Olofsson11}. We adopt the stellar photospheric spectrum from the 5600~K\footnote{We note that the temperature of young stellar objects can be influenced by many factors. We tried with a temperature range of 5100-5900K, and it doesn't affect the main points discussed in this paper.} \citep[e.g.,][]{Brown07}{}{} BT-Settl (AGSS2009) \citep[e.g.][]{AllardBTSettl1,AllardBTSettl2}{}{} with the assumption of solar metallicity. The accretion shock term is hard to determine for T~Cha due to the lack of contemporary FUV data, and has little impact on the mid-IR modelling and our main results. Thus, we ignore this term for simplicity. To run RADMC-3D we smooth the input stellar photospheric model with scipy.interpolate.interp1d and scipy.signal.savgol\_filter to remove absorption lines, making sure that the bolometric luminosity is conserved.
%As the model input, the stellar spectrum input is from the BT-Settl (AGSS2009) photospheric model  \citep[e.g.][]{AllardBTSettl1,AllardBTSettl2}{}{} with \textbf{$R_*=1.3R_\odot$, $M_*=1.5M_\odot$,} a solar metallicity, and a temperature of 5600~K\footnote{We note that the temperature of young stellar objects can be influenced by many factors. We tried with a temperature range of 5100-5900K, and it doesn't affect the main points discussed in this paper.}. Then we smooth the input stellar photospheric model with scipy.interpolate.interp1d and scipy.signal.savgol\_filter to get rid of absorption lines. 
We use the dsharp\_opac package from \cite{Birnstiel18} to generate the input opacity file of the dust incorporating constants from \cite{Draine03} for graphite and \cite{Weingartner01} for astronomical silicate. 
With the input stellar photosphere and the disk model with opacity file, the code performs radiative transfer calculations to generate the temperature structure and the SED. 
The best-fit model is found by minimizing the $\chi^2$ between the observed mid-IR SED and that derived from our model.

\subsubsection{Disk structure}
We parameterize both the inner and outer disks as follows: 
for both the inner and outer disk, we assume a vertical mass distribution that follows a Gaussian distribution (exp[$-\frac{z^2}{2H(r)^2}$]) at each radius $r$ \citep[e.g.,][]{Dullemond04}{}{}, where $H(r)$ is the disk scale height $H(r)=H_0(r/r_0)^\beta$ with a flaring exponent $\beta$ \citep[e.g.,][]{Dullemond04}{}{}. 
We also assume the surface density profile of a viscously evolving disk $\Sigma(r)= \Sigma_0 (r/r_0)^\alpha \exp{[-(r/r_{tap})^{2-\alpha}]}$ with an exponent $\alpha$ \citep[e.g.,][]{Hartmann98,Deng23}{}{}. In addition, each disk model is also characterized by an inner and outer radius and a total dust mass. For each disk component, the inner edge is a sharp cut at $r_{\rm in}$ while for the outer edge we apply an exponential decay where the outer radius $r_{\rm out}$ is set as the tapering radius $r_{tap}$ of it.
%where the $r_{\rm out}$ set as the $r_{tap}$ in the surface density profile equation. 
The dust properties are set by the minimum and maximum grain size ([$a_{min}$,$a_{max}$]) with a powerlaw distribution \citep[$n(a)\propto a^{-p}$, e.g.,][]{MRN77,Weingartner01}{}{}. 

\begin{figure*}[htb!]
    \centering
	\includegraphics[width=0.99\textwidth]{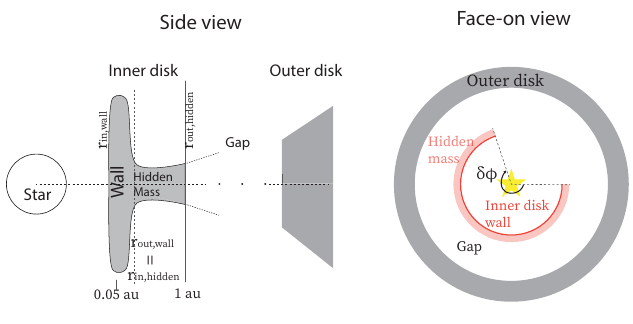}
    \caption{Sketch of the dust disk structure of our model. Sizes are not to scale. We fit three components: the inner disk dust wall, the hidden mass behind the wall and the outer disk together. The inner disk components can be azimuthally asymmetric.
    }
    \label{fig:sketch1}
\end{figure*}

Before the JWST observation, several studies have focused on modelling the SED of T~Cha and have provided good constraints on the outer disk \citep[e.g.,][]{Brown07,Olofsson13,Huelamo15}{}{}. However, previous models have not taken the optical variability into account and assume an azimuthally symmetric inner disk. Inspired by the hypothesis that the inner disk blocking the stellar photosphere is the cause of short-term optical variability \citep[e.g.,][]{Covino92,Herbst94}{}{}, we adopted an asymmetric inner disk model to fit the {\it Spitzer} spectrum along with the optical variability. We consider that all the mass in the inner disk is concentrated in $\delta \phi$ by setting the mass in other regions to zero (see Fig~\ref{fig:sketch1} for an illustration), and ignore the effect of dynamics caused by heating and cooling. In this way, when the optically thick part rotates between the star and our line of sight, we will see a highly extincted stellar photosphere, and when it rotates away there will be little/no extinction. On the other hand, for the less variable ($\delta V < 0.2$ mag in most cases) bright optical points between 2021 and 2023, extinction from the inner disk wall is not necessary. For simplicity, we use an azimuthally symmetric inner disk to model the SED at the time of the JWST observations.
We note that the $\delta \phi$ also represents an azimuthal concentration of inner disk wall and can significantly influence the SED, thus it is highly degenerate with other parameters in SED fitting (see appendix~\ref{app:connection} for more details). As such, we sample $\delta \phi$ coarsely (from $0^\circ$ to $360^\circ$ with a step of $60^\circ$) to simplify our modelling while showing the trend of how it influences both the optical and IR emissions.

We note that in the previous model for the {\it Spitzer} spectrum, all the shorter wavelength emission ($\sim$3-10$\mu m$) is from the very close-in part of the inner disk (i.e., the inner disk wall). A significant amount of dust mass (up to $\sim 5\times 10^{-11} M_\odot$) can be shadowed and hidden behind the optically thick inner disk wall that do not emit at mid-IR bands \citep[e.g.,][]{Olofsson13}{}{}. Thus, as illustrated in Figure~\ref{fig:sketch1}, we fit three components in our model, including an inner disk wall, the hidden mass in the inner disk behind the wall, and the outer disk. %In this model, the hidden component has a very low scale height compared with the wall, thus the area 
The parameter samplings of the inner disk wall and the outer disk are shown in Table~\ref{tab:fixpara}. 

Since almost all of the disk emission comes from the inner wall and the outer disk \citep[e.g.,][and in our model discussed in the following sections]{Olofsson13}{}{}, we add the hidden mass component only for estimating the upper limit of the micron-sized dust mass in the inner disk, thus its azimuthal distribution also follows the inner disk wall.  The amount of mass in this component is an upper limit because if more mass is added it will no longer be hidden and contribute to the mid-IR emission resulting in a silicate emission feature at $\sim 10$\,micron, which is not consistent with the observations.
The ALMA image constrains the inner disk of T~Cha to be smaller than 1~au \citep{Hendler18}, which we set as the outer radius of the hidden mass.
%thus we set the inner radius of the hidden mass ($r_{\rm in, hidden}$) as the outer radius of the inner disk wall ($r_{\rm out, wall}$), and the outer radius of the hidden mass ($r_{\rm in, hidden}$) as 1~au. \cyxie{some re-phrasing} 
Other parameters of the hidden mass are set or sampled the same as the inner disk wall, with the scale height limited to be lower than the wall. 
How the fixed parameters in Table~\ref{tab:fixpara} are chosen is discussed in detail in the following section~\ref{model:input}.

\subsubsection{Input parameters} \label{model:input}
Following previous studies mentioned above, and assuming the outer disk has not changed during previous decades, we take the value of inclination, PA, $\alpha$, $\beta$ and outer radius of the outer disk from \cite{Pohl17} (see table~\ref{tab:fixpara}), which is also consistent with recent analysis of ALMA high-resolution continuum data \citep{Hendler18}. 
%\sout{ We note that in the model of \cite{Espaillat11}, the change in the outer disk emission is from blocking of the inner disk wall, thus we assume an azimuthally symmetric outer disk well aligned with the inner disk. } \ilaria{The sentence above breaks the logic, should go somewhere else}
The inner edge of the inner disk is set as 0.04~au ($\sim$7$\times$R$_\odot$), 
%near the co-rotational radius $R_{\rm co}$\citep{McGinnis15} 
%\ilaria{did you calculate the co-rotaion radius or did McGinnis did for TCha? or is this a general statement that the co-rotation is typically at 5Rstar? Clarify} \cyxie{it seems that in McGinnis the Rco is not calculated, but measured from some observations. I took 5Rstar for most of the paper take 5Rstar also as the Rmag(truncation radius), \citep[e.g.,][]{Gullbring98,McGinnis15}{}{} and also Tsang+11}
very close to the radius corresponding to the $\sim 3.3$ days optical variability. Note that this is a reasonable value given that the magnetic truncation radius is expected to between 0.03-0.06~au (changing the inner edge between these values does not influence the main results, see app.~\ref{app:connection}).  With a number density of $\sim 10^{10}$ cm$^{-3}$ in the inner disk wall and equations~32 and 33 in \cite{Baskin18}, the sublimation temperature of astronomical silicates is $\sim$1500~K and that of graphite is $\sim$1900~K. Based on the stellar properties,  the dust temperature at the inner disk edge is $\sim$ 1800~K, meaning that silicates are sublimated and unlikely to be located as close as 0.04\,au while graphite grains can be present.

%\ilaria{who said that? Did you use RADMC3D to calculate it? or are you setting it up to be 1800K} \cyxie{this is from Radmc-3D modelling}
%\textbf{This value is higher than the sublimation temperature of astronomical silicate ($\sim$1500~K) but lower than that of graphite ($\sim$2000~K), see \cite{Baskin18}\footnote{\cyxie{With an estimation of the number density $\sim 10^{10}$ cm$^{-3}$ in the inner disk wall}}, meaning that silicates are unlikely to be located as close as 0.04\,au while graphite grains can be present.}
Thus, we set the dust species of the inner disk wall as purely graphite, while we adopt a dust composition consisting of 70\% astronomical silicates and 30\% graphite by mass fraction \citep[e.g.,][]{Weingartner01}{}{} for the outer disk and the inner disk hidden mass behind the wall. 
Here we note that the silicates can still be present in the inner disk wall but need to be a minor component as no  silicate feature is detected in the Spitzer or JWST spectra.
We also assume an azimuthally symmetric outer disk well aligned with the inner disk for both {\it Spitzer} and JWST models.

During our modelling, we find that after fixing the dust composition, the slope of the SED at far-infrared and sub-millimeter wavelengths only depends on the mass and grain size distribution of the outer disk. %, the very low-mass inner disk doesn't influence that range. 
Thus, we first fit the {\it Herschel}, SEST and ATCA photometry at wavelengths longer than 100$\mu$m \citep{Lommen07,Cieza11,Ubach12}, to estimate the dust mass $M_{\rm dust, out}=1\times10^{-5} M_\odot$ and the index of the power law dust distribution $p=3.7$, which are kept fixed subsequently.

The inner radius of the outer disk ($r_{\rm in, out}$) has been constrained via NIR interferometry, SED fitting \citep{Olofsson11,Olofsson13}, near-IR scattered light \citep{Pohl17} and millimeter ALMA continuum data \citep{Hendler18}, and resulting in estimates ranging from 12~au to 30~au. With our model, we found that the inner radius of the outer disk is a crucial parameter and needs to be closer than 24~au so that the outer disk can become hot enough to emit sufficient flux at 15-25\,\micron\ even if the inner disk is not blocking any stellar radiation to the outer disk. We allow this parameter to change during our fitting process, and keep it the same between the {\it Spitzer} and JWST models. The sampling of other parameters is shown in Table~\ref{tab:fixpara}. 
Some of the parameters of the inner disk wall, including the $\alpha$, $\beta$, and width ($r_{\rm out,wall}-r_{\rm in,wall}$) do not influence the SED fitting (see appendix~\ref{app:connection} for a detailed discussion). The dust grain size distribution, on the other hand, is highly degenerate with the dust mass because only the small micron-sized grains contribute to the mid-IR SED emission. Thus we take $\alpha$, $\beta$, and the grain size distribution of the inner disk wall the same as \cite{Pohl17} and keep the $r_{\rm out,wall} =0.06$\,au. 
How the SED changes with each parameter is further discussed in Appendix~\ref{app:connection}.

%We note that in the model for the {\it Spitzer} spectrum, all the shorter wavelength emission ($\sim$3-10$\mu m$) is from the very close-in part of the inner disk (i.e., the inner disk wall). A large amount of dust mass (up to $\sim 5\times 10^{-11} M_\odot$) can be hidden behind the optically thick inner disk wall \citep[e.g.,][]{Olofsson13}{}{}. Thus, the parameters in Table~\ref{tab:fixpara} only account for the inner disk wall. In our model, we also add another component within 1~au behind the inner disk wall to estimate the upper limit of the mass of micron-sized grains in the inner disk. 

\begin{deluxetable*}{c|cc}
%\tablenum{2}
\tablecaption{Sampling of parameters\label{tab:fixpara}}
\tablewidth{1.2\textwidth}
\tablehead{
%\multicolumn{3}{c}{Fixed parameters}\\
%\hline
Parameters & Inner disk wall& Outer disk\\
%& $r_{in}$[au] & $r_{out}$[au] &$r_c$[au]& $M_{dust}$[$M_\odot$] & $\alpha$ & $\beta$ &$H_0/r_0$ & $p$ & $a_{min}$[$\mu m$] & $a_{max}$[$\mu m$] & incl[$^\circ$] & PA[$^\circ$] \\
}
\startdata
$r_{\rm in}$[au] &0.04\tablenotemark{1}& [12, \textbf{15}, 18, 21, 24] \\
$r_{\rm out}$[au]%\tablenotemark{2} 
&[\textbf{0.06}, 0.08, 0.1, 0.12, 0.14, 0.16]& 50\tablenotemark{a}\\
$M_{\rm dust}$[$M_\odot$]  & From $10^{-10}$ to $10^{-13}$, with 40 steps in log space& [0.1, 0.5, \textbf{1}, 5, 10]$\times$10$^{-5}$ \\
$\alpha$ & -1 & -1 \\
$\beta$ & 1 & 1 \\
$H_0/r_0$ & [0.004, ..., 0.04]/0.1, with a step of 0.004 &[1.5, 2.0, \textbf{2.5}]/25\\
\hline
$p$ &3.5\tablenotemark{a} & [3.3, 3.4, 3.5, 3.6, \textbf{3.7}, 3.8]\\
$a_{\rm min}$[$\mu m$] & \multicolumn{2}{c}{0.01} \\
$a_{\rm max}$[$\mu m$] & \multicolumn{2}{c}{1000} \\
\hline
$i$[$^{\circ}$] & \multicolumn{2}{c}{69\tablenotemark{b}} \\
$PA$[$^{\circ}$] & \multicolumn{2}{c}{114\tablenotemark{b}} \\
%Inner disk& 0.04\tablenotemark{1} & 0.06 & -& - & -1 & 1&- & 3.5& 0.01 &  \multirow{2}{*}{1000} & \multirow{2}{*}{69\tablenotemark{b}} & \multirow{2}{*}{114\tablenotemark{b}}\\
%Outer disk& - & 60\tablenotemark{b} & 50\tablenotemark{a} & \underline{$1 \times 10^{-5}$}& -1& 1& \underline{2.5/25} & \underline{3.7} & 0.01  &  & &\\
\enddata
\tablecomments{Values shown in boldface are first fit and fixed then between both {\it Spitzer} and JWST models. 
\tablenotetext{1}{Estimated with the period of optical variability and the co-rotational radius.}
%\tablenotetext{2}{ which does not significantly influence the SED (see app~\ref{app:connection} for more details).} 
To estimate the upper limit of the mass in the inner disk, the $r_{\rm hidden, out}$ is set at 1~au.}
\textbf{References:} \tablenotemark{a} Huelamo et al. (2015); \tablenotemark{b} Pohl et al. (2017)
\end{deluxetable*}

\subsection{Fitted models}
%\subsection{Symmetric model}
%T~Cha has been modelled many times before, and the {\it Spitzer} spectrum of it was already well fitted \citep[e.g.,][]{Brown07,Olofsson13,Huelamo15}{}{}. 

\begin{figure*}[htb!]
    \centering
	\includegraphics[width=0.99\textwidth]{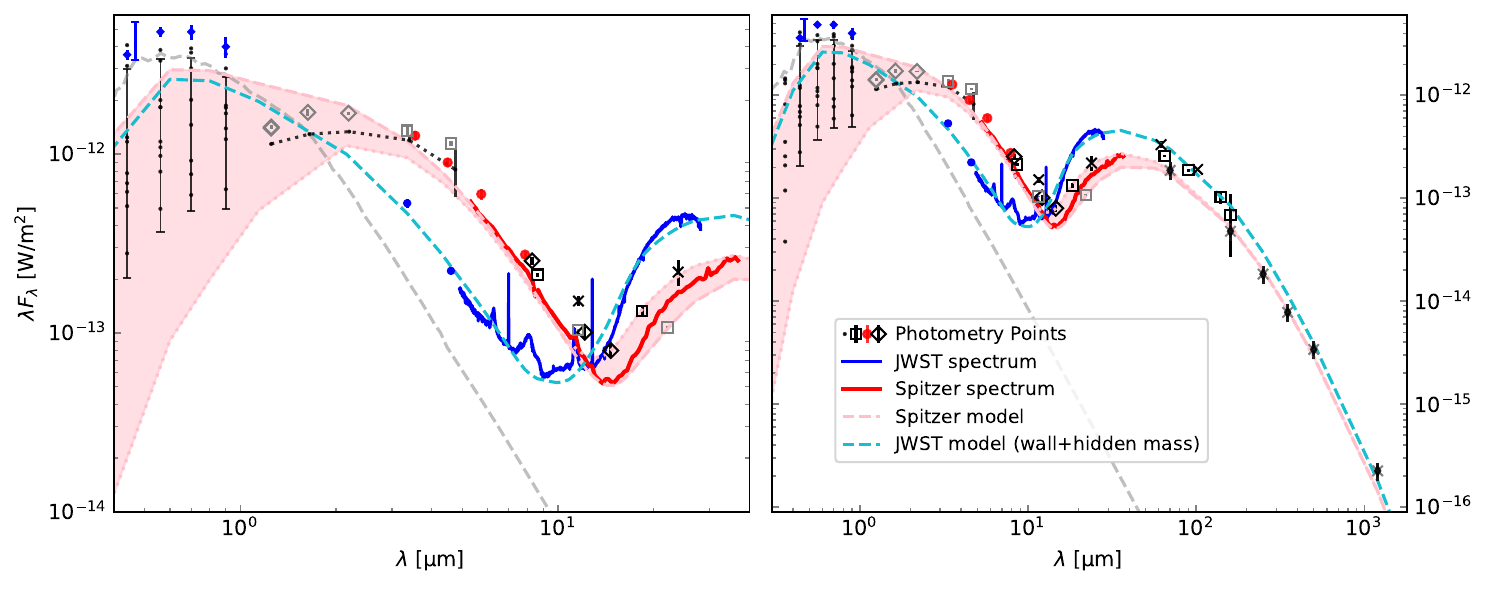}
    \caption{Comparison between our fitted model with the observed spectra. Left panel is the zoom-in version between $0.4$ and $40\mu m$ and the right panel is the full SED. The photometric points, observed spectra and photosphere model are the same as Fig.~\ref{fig:Overview}. The pink and cyan dashed and dotted lines are our model for {\it Spitzer} and JWST spectra, respectively. The pink shaded region indicates the maximum optical mid-IR variability that can be caused by the azimutally asymmetric inner disk, with the dashed line indicating no-extinction case and dotted line indicating extinction case. 
    }
    \label{fig:asym}
\end{figure*}

By minimizing the chi-square between our model at mid-IR wavelengths (5-25 $\mu m$) with the spectra from either {\it Spitzer} or JWST, we find a best-fit model for each of them.
The results of our best-fit models are summarized in Table~\ref{tab:fittedpara}, while the SEDs compared to the observations are shown in Fig~\ref{fig:asym}. The observed photometric points, spectrum and the input stellar SED are also plotted in Fig~\ref{fig:Overview}. 
The shaded region of the model for the {\it Spitzer} spectrum indicates the model with the same parameters but the optically thick part of the inner disk at different azimuthal locations. The dotted pink line indicates the optically thick part is between the star and us, while the dashed line indicate it is away from us. The different azimuthal location of the optically thick wall will also cause a slight difference at mid-IR wavelengths (shaded region in Fig.~\ref{fig:asym}). 
%
%With an inclination of $\sim 70^\circ$, a small part of the inner and outer disk ($\sim 30^\circ$) between the star and our line-of-sight will be mostly self-blocked (both inner and outer disk) and will cause a slight difference between different phase of the azimuthal location of the optically thick part (e.g., fig~\ref{fig:asym}). 

%To reproduce the Spitzer data, the scale height of the inner disk $H/r=0.36$ needs to be much larger than in hydrostatic equilibrium case ($H/r\sim0.02$). This high scale height can be caused by the coupling between gas and dust in the inner hot part of the disk \citep{Thi11}. From our model, we find decreasing the inner disk wall height can reproduce the seesaw effect while accounting for the optical photometry behavior. 
\begin{deluxetable*}{c|c|cc}
%\tablenum{4}
\tablecaption{Fitted models\label{tab:fittedpara}}
\tablewidth{1.2\textwidth}
\tablehead{
%\multirow{2}{6em}{Parameter} & \multicolumn{2}{c}{\multirow{3}{*}{Symmetric}} &\multicolumn{6}{c}{Asymmetric}  \\
\multirow{2}{6em}{Parameter} & {\it Spitzer} & \multicolumn{2}{c}{JWST} \\
& Asymmetric &  \multicolumn{2}{c}{Symmetric} \\
& Wall & Wall & Hidden mass$^*$  \\
}
\startdata
$(r_{\rm in}, r_{\rm out}) [au]$& (0.04, 0.06)& (0.04, 0.06)& (0.06, 1)\\
$M_{\rm dust,in}$[$\times$10$^{-12}M_\odot$]& 7.0 &0.8 & 35 \\
$(H_{0}/r_{0})_{\rm in}$ (at $r_{0}=$ 0.05~au)& 0.36 & 0.08& 0.04\\
$\delta \phi_{\rm in}$ [$^\circ$]&  300 &\multicolumn{2}{c}{-}\\
\enddata
\tablecomments{
Fitted models for both {\it Spitzer} and JWST SED. The decrease of the inner disk wall mass or height can result in the mid-IR seesaw behavior between the SEDs. \\
$^*$ This is an estimation on the upper limit of the micron-sized dust mass that can be hidden behind the inner disk wall within 1~au. 
}
\end{deluxetable*}
We note that our asymmetric model is an extreme case with the assumption that all the inner disk mass is azimuthally concentrated in $\delta \phi$ and the whole disk is always in thermal equilibrium. The real mass distribution can be more uniform, i.e. there can be some optically thin or low scale height dust disk in the region where we set the dust mass equal to zero contributing to less variable optical extinction than what is assumed in our model. In addition, the cooling timescale of the outer disk \citep[$T_{\rm cool}>0.5$ year, estimated from equation.~39 of][ with their assumptions and an outer disk inner edge temperature $T\sim 100$K]{Zhang20}{}{} is much longer than the rotational period of the inner disk ($\sim 3$ days), indicating the outer disk will react less efficiently to the azimuthal location of the inner disk.
Thus, the shaded region at both the optical and longer wavelengths ($>10 \mu m$) should be interpreted as the maximum variability that can be caused by the inner disk rotation.

%The real extinction caused by the inner disk should be less than shown in the figure since the real mass distribution \textbf{can be more uniform}, so is the outer disk SED considering the cooling timescale of the outer disk \citep[$T_{\rm cool}>0.5$ year, estimated from the equation.39 of][ with their assumptions and an outer disk inner edge temperature $T\sim 100K$]{Zhang20}{}{} is much longer than the rotational period of the inner disk ($\sim 3$ days).
%Thus, the shaded region in Fig.~\ref{fig:asym} should be interpreted as the maximum variability that can be caused by the inner disk rotation (for both optical and mid-IR wavelengths). 

As discussed in Section~\ref{sec:model} and \cite{Olofsson13}, a large amount of mass can be hidden behind the optically thick inner disk wall. To estimate the upper limit of the mass, we added another component of dust mass behind the inner disk wall with a lower scale height. Considering the inner disk wall height is much lower at the JWST time, hence the maximum hidden mass can be better constrained, we only show the fitted parameters of the additional component in our JWST model. 
%In addition, it is possible that the inner disk wall during the JWST time is sublimated so much that it cannot act as a ``wall'' that casts enough shadow on the previously hidden mass (see Fig~\ref{fig:sketch2} for a sketch). 
%In this case we apply an inner disk with a $r_{\rm in}=0.04$au, $r_{\rm out}=1$au, and other parameters sampled as Table.~\ref{tab:fixpara} to fit the JWST spectrum as well. 
With a good fit to the JWST spectrum, we find an upper limit to the hidden mass in the inner disk of $\sim 3.6\times 10^{-11} M_\odot$, consistent with what is reported in \cite{Olofsson13} (our value is slightly lower than their estimate $M_{\rm hidden} \sim 5\times 10^{-11} M_\odot$). The best-fit parameters of this model are reported in Table~\ref{tab:fittedpara}. Besides this, we also explored the possibility that the inner disk wall is completely sublimated and the previously hidden mass is the main contribution to the NIR emission. However, in this case there will be a strong 10~$\mu$m silicate feature, which is not observed in our JWST spectrum (see appendix~\ref{app:connection} for more details). Thus, we can rule out the possibility of complete destruction of the inner wall.

\begin{figure*}[htb!]
    \centering
	\includegraphics[width=0.40\textwidth]{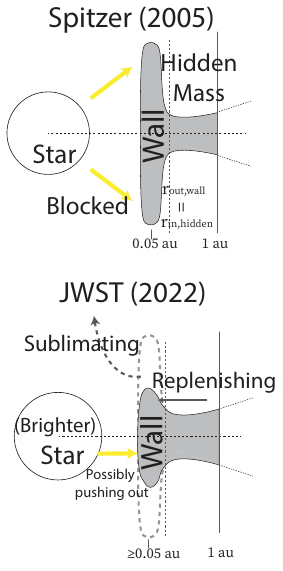}
    \caption{Sketch of our hypothesis about how the disk changes from {\it Spitzer} to JWST time. When the star gets brighter and sublimates the dust grains in the inner disk wall, the wall reduces in height significantly, blocking less radiation to the outer disk and lead to less optical extinction. The replenishment from the hidden mass is either enough to still maintain a lower wall (wall $+$ hidden mass model) or not (disk without a wall model).
    }
    \label{fig:sketch2}
\end{figure*}

We find that an azimuthally asymmetric inner disk wall can reproduce the {\it Spitzer} spectrum and the strong optical short-term variability of T~Cha at the same time. The strong mid-IR seesaw behavior seen in the JWST spectrum along with the higher and less variable optical emission between 2021 and 2023 can be caused by a significant decrease of the inner disk wall height. 
Due to the degeneracy between parameters, there are also some other models with a slightly higher chi-square that can reproduce the data, but all of them follow the same trend as discussed above.
Thus we propose that one scenario to explain the observations is that in 2021 an outburst or change in stellar magnetic field caused the inner disk wall to significantly decrease resulting in the seesaw behavior between the {\it Spitzer} and JWST spectra, along with the steady high optical emission (see Figure~\ref{fig:sketch2} for an illustration). %
%\sout{\cyxie{Although a stellar outburst would lead to higher stellar luminosity and push the sublimation radius outwards, it has a different impact on the surface and the mid-plane because the dust sublimation temperature is correlated with  the density \citep{Baskin18}. Based on our model, the dust density $n$ in the mid-plane is $\sim 10^{11} cm^{-3}$, corresponding to a $T_{sub} \sim 2000$~K, and decrease with scale height following a gaussian distribution to $n \sim 10^9 cm^{-3}$, corresponding to a $T_{sub} \sim 1800$~K. In this way, if the stellar luminosity doubles, it can sublimate a large fraction of the inner wall, but only push the mid-plane sublimation front to $\sim 0.045$~au, hence our disregard of this outward movement in our analysis.}} 
Because the dust sublimation temperature increases with dust density \citep{Baskin18} and the density of small grains follows the gas density, which has a vertical Gaussian profile, dust  in the midplane is less prone to sublimate 
(while dust above and below sublimates more easily). Doubling the stellar luminosity during the 2021 outburst would only shift the midplane sublimation front to $\sim0.045$\,au, which is why we neglect the outward movement of the inner wall in our analysis.
%\textbf{Although a stellar outburst would push the inner edge of the wall outward, the increase in luminosity would only shift the wall to 0.56\,au (from 0.04\,au), which does not significantly affect the SED (see Appendix~\ref{app:connection}), hence our disregard of this outward movement in our analysis.}
%\textbf{We also note that the brighter star can push the inner edge of the inner disk outwards (e.g., when $L_*$ doubles, the inner disk edge will increase by $\sim 41 \%$ from 0.04 to 0.56~au), but that would not influence the SED modelling a lot (see app.~\ref{app:connection} for a detailed discussion). Thus, we ignore this effect in our modelling.}
The highly variable optical data and increasing WISE photometry past 2023 suggest that the asymmetric inner disk wall present during the Spitzer observations has built up again.

%\textbf{We note that the stellar luminosity during this period is $\sim 2$ times brighter than our 5600~K photospheric model (see the difference between blue diamonds and gray dashed line in Fig~\ref{fig:asym}). If all this brightening comes from less extinction from the disk, T~Cha should be as hot as $>6500$~K, indicating it's spectral type is earlier than F5 \citep[e.g.,][]{Pecaut13}{}{}. This is significantly different from previous studies suggesting a non-variable G8 type \citep[e.g.,][]{Alcala93,Schisano09}{}{}, implying this optical brightening comes from an outburst. }

%%Small grain mass decreased

\section{Discussion}\label{sec:dis}

\subsection{Explanation of our model}
%\subsubsection{Physical causes of the asymmetric inner disk wall}
In the previous sections, we find that the short-term optical variability seen in T~Cha can be caused by an asymmetric inner disk wall while the significant mid-IR seesaw behavior between {\it Spitzer} and JWST spectra can be caused by the inner disk wall decreasing in height. 
%\ilaria{WHY DO WE NEED THE NEXT SENTENCE? I WOULD REMOVE IT, WE SAY THAT ANYWAY IN THE NEXT PARAGRAPH} \textbf{We note that our simple model  aims at providing an explanation of the optical and IR variability for T~Cha and try to explore the possible mechanisms caused that, instead of accurately constrain the parameters for the disk.}

Our model with all the dust mass concentrated in a certain azimuthal region is highly simplified, but a somewhat similar asymmetric inner disk wall can be caused by several mechanisms. At the very inner edge of disks, the dynamical interaction between the magnetic field and the disk can lead to the development of a disk inner wall with azimuthally varying height \citep[e.g.,][]{Bouvier99}{}{}. 
In addition, material from the disk accretes onto the star following the stellar magnetic field lines \citep[e.g.,][]{Hartmann16}{}{}. 
If the accretion originates outside the sublimation radius, the accretion flow can be dusty and thus cause NIR excess while occulting the star periodically \citep[e.g.,][]{Bovier07a,Gaidos24}{}{}.
Both the azimuthally varying disk wall and the dusty accretion flows can act as an azimuthally asymmetric inner disk wall and result in the optical variability (both the periodic one and the significant scatter on top) we have seen for T~Cha. 
In addition, the perturbation from the gravity of a planet located near the inner disk can also influence the disk dust mass distribution \citep[e.g.,][]{Benisty23}{}{}. 

We propose that the strong seesaw behavior between the {\it Spitzer} and JWST spectra along with the stronger but less variable optical emission between 2021 and 2023 is caused by the significant decrease of the inner disk wall height or mass. 
Recently, \cite{Espaillat24} analyzed the UX~Tau A transition disk showing very similar  seesaw behavior at mid-IR wavelengths. The shorter wavelength emission in their JWST spectrum reaches almost the photosphere level while longer wavelength emission increased compared with the {\it Spitzer} spectra $\sim 15$ years ago. The contemporary optical photometry is also brighter and less variable. \cite{Espaillat24} argued that the precession of the inner disk leads to misalignment with the outer disk, disrupting the replenishment of the inner disk. This, along with the constant accretion from the inner disk to the star, depleted the inner disk of micron-sized dust grains, leading to less emission from the inner disk and less shading of the star and the outer disk. 

T~Cha is  similar to UX~Tau A except that at the JWST time T~Cha has significantly more NIR and optical emission  compared with the photosphere emission of a G8 type star even without de-extincting the data. Thus, depleting the whole inner disk is not compatible with the data available for  T~Cha. We argue, instead that the seesaw behavior of the T~Cha spectrum is likely due to a stellar outburst happening between 2021-2023. This explanation agrees with the brightening at optical wavelengths and the stronger ionic lines and PAH features in that period \citep[see Fig~\ref{fig:Photometry},~\ref{fig:AAVSO} and ~\ref{fig:Overview} for a reference and][for more details]{Bajaj24}. For the brighter optical emission to be explained simply by less extinction from the inner disk wall, T~Cha would require a temperature of at least 6000~K, corresponding to a spectral type earlier than G0 \citep[e.g.,][]{Pecaut13}{}{}. This is significantly different from previous spectra of T~Cha showing that it is consistently a G8 type star \citep[e.g.,][]{Alcala93,Schisano09}{}{}. The discrepancy implies this optical brightening can come from an outburst, and can be tested with future optical spectroscopy targeting gas lines.

When the stellar luminosity increases (during the outburst phase), the dust temperature at the inner disk wall increases from $\sim$1800\,K  to $\sim$2160\,K and even graphite grains sublimate.  In this way, the stellar outburst could have reduced the wall height significantly, and the strong stellar emission during this period prevents a new tall dust wall to be built. 
In our model, the innermost asymmetric part has been mostly sublimated and will contribute less to the continuum emission, hence we use a symmetric wall with reduced height and mass (see section~\ref{sec:model}). 
After the outburst 
%\ilaria{NOTE: THE STAR HAS MORE VARIABILITY SO I WOULD NOT USE THE WORD QUIESCENT UNLESS WHEN WE REALLY HAVE TO, NEXT PARAGRAPH. THE OTHER THING IS THAT OUTBURSTS OF 2 YEARS DO NOT EXIST IN NATURE AS FAR AS I KNOW SO I WOULD NOT CLAIM ANYWHERE THAT THE OUTBURST IS 2 YEARS MAYBE IT JUST TOOK 2 YEARS TO REBUILD THE WALL}, 
the hidden dust mass moves inward and starts to build up the tall inner disk wall again. 

Such stellar outbursts can be caused by several mechanisms, including the instability of the inner disk edge near the co-rotation radius, and reversal of the stellar magnetic field driven by dynamos \citep{DAngelo10,Armitage16,Fischer23}. The former scenario has been discussed in detail by \cite{DAngelo10} and could explain our data as follows. At the quiescent stage, the inner edge of the gas disk (which is at the magnetic truncation radius) is slightly outside the co-rotation radius meaning that the star rotates slightly faster than the inner disk. The coupling between the stellar magnetic field and the disk material accelerates the disk material, preventing it from moving inward, suppressing the accretion, and possibly driving outflows \citep[e.g.,][]{Ustyugova06}. During the quiescent stage, the sublimation radius would be at or inside the truncation radius, both gas and dust would pile up in the inner region of the disk, building up a tall dust wall at the truncation radius. The inner edge slowly moves inwards as mass piles up, eventually crosses the co-rotation radius. After that, the accretion increases significantly  and empties the whole inner region during a short period, causing an accretion outburst while pushing the inner edge of the dust disk outward. During this outburst phase, both the corotation radius and the magnetic truncation radius would be inside the sublimation radius $-$ the dust in the disk will thus cross this radius and  sublimates, forming only a lower inner wall at that location due to the pressure maximum by dust sublimation. When the outburst ends, the location of the sublimation radius moves closer to the star, and the system is back to a quiescent stage again. If this scenario is correct, we can also set a constraint on the stellar period of T~Cha as shorter than or equal to the $\sim 3.3$ days period as that would set the corotation radius.

\subsection{Implications of our Model and Future observations}
In the previous sections, we proposed that both the optical and mid-IR SED changes are caused by a stellar outburst sublimating and destroying most of the inner disk wall. Similar brightening of optical emission was reported in 1989 \citep{Alcala93}, 2006, and 2015 (see the blue shaded area of Fig.~\ref{fig:Photometry}), suggesting a recurring event every 5-10 years. This episodic accretion outburst agrees well with the model discussed in \cite{DAngelo10}, and provides several distinct implications. 

In the model from \cite{DAngelo10}, during the outburst phase, all the dust mass at the innermost edge of the disk will be sublimated and accreted onto the star in a short period. Based on our model, that would imply that a mass of at least $\sim 7\times 10^{-12}M_\odot$ in small grains is removed in 1-2 years. Previous studies \citep[e.g.,][]{Olofsson13} as well as our model suggest that the mass of micron-sized grains in the whole inner disk is less than $3.5 \times 10^{-11}M_\odot$. On the other hand, with the millimeter flux from \cite{Hendler18} and subtracting the free-free emission \citep[e.g.,][]{Pascucci14,Rota24}, the 3 mm emission from the dust ($\sim 0.1$ mJy) gives a total inner disk  mass of $\sim 10^{-8} M_\odot$. The relatively high millimeter dust mass suggests a significantly greater abundance of millimeter-sized grains compared with micron-sized grains. This is also in agreement with the high extinction ratio $R_V \sim 5.5$ \citep[e.g.,][]{Covino97, Schisano09} from the disk compared with the average value ($R_V =3.1$) for the ISM, which indicates probable grain growth and depletion of small grains \citep{Draine09}.

%\sout{\cyxie{Maybe remove this part}
%From these estimations, \textbf{if there is no other source replenishing the inner disk mass,} we will be able to see the inner disk of T~Cha deplete most of its micron-sized dust grains in decades (after $\sim$5 outburst events) which will lead to a significant difference in the SED. }
The absence of a gap in the gas disk, along with accretion, suggest that micron-sized grains from the outer disk, which are coupled with the gas, will likely replenish the inner disk. Future ALMA observations and modeling might constrain this replenish rate, providing better insights into on how long the inner  micron-sized dust will take to be depleted. 
Additionally, high spatial resolution instruments like VLT/MATISSE can spatially resolve the inner disk of T~Cha, which would allow  to further analyze the differences between its outburst and quiescent stages. High-resolution UV or optical spectra taken during the next  optical bright and relatively constant stage could constrain the accretion rate at that time, verifying whether T~Cha is experiencing an accretion outburst. Future JWST observations in both quiescent and outburst stages would also reveal whether the spectra are similar to previous observations, while also exploring any potential effect from disk winds \citep[e.g.,][]{Bajaj24,Sellek24}{}{}.

\section{Summary}\label{sec:sum} 
%\begin{itemize}
%    \item We presented the new observation of TCha with JWST and discovered a see-saw behavior which is different from all previous mid-IR observations.
%    \item The behavior along with the optical variability can be explained by an asymmetric inner disk and an evolution of the inner disk during 17 years.
%    \item This suggests the inner disk of TCha is at the last stage of its evolution and will disperse and deplete in several years. 
%\end{itemize}

In this study, we have acquired and analyzed the JWST/MIRI-MRS spectrum of T~Cha, an accreting G8 star surrounded by a transition disk, exhibiting highly variable optical emission. This disk is known for its large dust gap separating the small inner disk ($<$1~au) and the outer disk ($\sim$ 20-50~au). The MIRI-MRS spectrum of T~Cha shows an unexpected change (seesaw variability) compared to a previous {\it Spitzer} spectrum and all other photometric points observed before 2020. The emission at shorter wavelength ($\lambda <10\mu m$) decreases while at longer wavelength ($\lambda >10\mu m$) increases, both by a factor of up to 3. 
This mid-IR change is consistent with recent WISE observations and accompanied with a less variable but brighter optical emission during 2021-2023. 
We use the radiative transfer code RADMC-3D to model  the {\it Spitzer} and JWST spectra along with the entire SED, and find that both the mid-IR and optical behavior of T~Cha can be explained by an optically thick asymmetric inner disk wall with significantly decreased  mass and height, possibly due to an outburst.

Our observations and models suggest that T~Cha is  experiencing episodic events, possibly accretion outbursts, every 5-10 years. In each event, an appreciable amount of micron-sized dust grains are removed from the inner disk. This suggests that T~Cha may show a significant change in its SED in the following decades due to the decrease and depletion of the micron-sized dust grains, and will be an excellent system to study inner disk dust replenishment. Future high spatial and spectral resolution observations  at different bands can help understand the last stage of evolution of this and other transition disks. 
%Our observation and model suggest that the inner disk of T~Cha is at its last stage of evolution, and will disperse all its dust mass in the following decades. Follow-up observations of T~Cha at different bands are important not only in verifying our model, but also in studying and understanding the last stage of disk evolution. 

\begin{acknowledgments}
This work is based on observations made with the NASA/ ESA/CSA James Webb Space Telescope. The data were obtained from the Mikulski Archive for Space Telescopes at the Space Telescope Science Institute, which is operated by the Association of Universities for Research in Astronomy, Inc., under NASA contract NAS 5-03127 for JWST. Some/all of the data presented in this paper were obtained from the Mikulski Archive for Space Telescopes (MAST) at the Space Telescope Science Institute. The specific observations analyzed can be accessed via \dataset[doi:10.17909/dhmh-fx64]{https://doi.org/10.17909/dhmh-fx64}. The observations are associated with JWST GO Cycle 1 program ID 2260. This work was funded by NASA/STScI GO grant JWST-GO-02260.001. We thank Andras Gaspar and Jane Morrison for their efforts in acquiring and processing the data. This work was also supported by the NKFIH excellence grant TKP2021-NKTA-64. C.X. and I.P. acknowledge partial support by NASA under Agreement No. 80NSSC21K0593 for the program ``Alien Earths”. A.D.S. was supported by the European Union's Horizon 2020 research and innovation program under Marie Sklodowska-Curie grant agreement No. 823823 (DUSTBUSTERS). G.B. has received funding from the European Research Council (ERC) under the European Union's Horizon 2020 Framework Programme (grant agree- ment No. 853022, PEVAP). A.D.S. acknowledges the support of a Science and Technology Facilities Council (STFC) PhD studentship and funding from the European Research Council (ERC) under the European Union's Horizon 2020 research and innovation program (grant agreement No. 1010197S1 MOLDISK). R.A. acknowledges funding from the Science \& Technology Facilities Council (STFC) through Consolidated Grant ST/W000857/1. C.X. thanks Chia-Lung Lin for the useful discussion on the variability mechanisms of variable stars.
We acknowledge with thanks the variable star observations from the AAVSO International Database contributed by observers worldwide and used in this research. This work is based in part on archival observations made with the Spitzer Space Telescope, which was operated by the Jet Propulsion Laboratory, California Institute of Technology, under a contract with NASA. 
\end{acknowledgments}

\vspace{5mm}
\facilities{JWST(MIRI MRS), {\it Spitzer}(IRS)}
%\facilities{HST(STIS), Swift(XRT and UVOT), AAVSO, CTIO:1.3m,


\software{{\tt astropy} \citep{2013A&A...558A..33A,2018AJ....156..123A,2022ApJ...935..167A}, {\tt JWST} \citep{Bushouse23}, {\tt scipy} \citep{Virtanen20} }



\appendix
\section{Connection between parameter changes and SED changes} \label{app:connection}
To model T~Cha, we have explored over ten parameters and find many of them are highly degenerate when it comes to SED fitting. In this section, we summarize how the SED changes with each parameter. We start with a model with a symmetric inner disk wall and other parameters similar to our {\it Spitzer} model as shown in Table~\ref{app:tab:Start}. In each model, we only change one parameter while others are kept constant. 
%We note that the outer disk parameters only influence the SED at longer wavelengths ($>10\mu m$), and has been well studied by \cite{Woitke16}. The hidden mass behind the inner disk wall does not influence the mid-IR SED. Thus, here we focus on changing the parameters of the inner disk wall. 

\begin{deluxetable*}{c|c|c}
%\tablenum{4}
\tablecaption{Starting model}\label{app:tab:Start}
\tablewidth{1\textwidth}
\tablehead{
%\multicolumn{3}{c}{Fixed parameters}\\
%\hline
Parameter & Inner disk wall & Outer disk\\
}
\startdata
$r_{\rm in}$[au] & 0.04 & 15\\
$r_{\rm out}$[au] & 0.06 & 50\\
$M_{\rm dust}$[$M_\odot$]  & $7\times 10^{-12}$ & $1\times 10^{-5}$\\
$\alpha$ & -1 & -1\\
$\beta$ & 1 & 1\\
$H_0/r_0$& 0.02/0.1 & 2.5/25 \\
$\delta \phi$ [$^\circ$]& 360& -\\
\hline
$p$ & 3.5 &3.7\\
$a_{\rm min}$[$\mu m$] & \multicolumn{2}{c}{0.01}\\
$a_{\rm max}$[$\mu m$] & \multicolumn{2}{c}{1000}\\
%Dust composition & & & &\\
\hline
$i$[$^{\circ}$] & \multicolumn{2}{c}{69} \\
$PA$[$^{\circ}$] & \multicolumn{2}{c}{114} \\
\enddata
\tablecomments{When one parameter changes, other parameters are kept constant as shown here. 
}
\end{deluxetable*}

\begin{figure*}[htb!]
    \centering
	\includegraphics[width=0.99\textwidth]{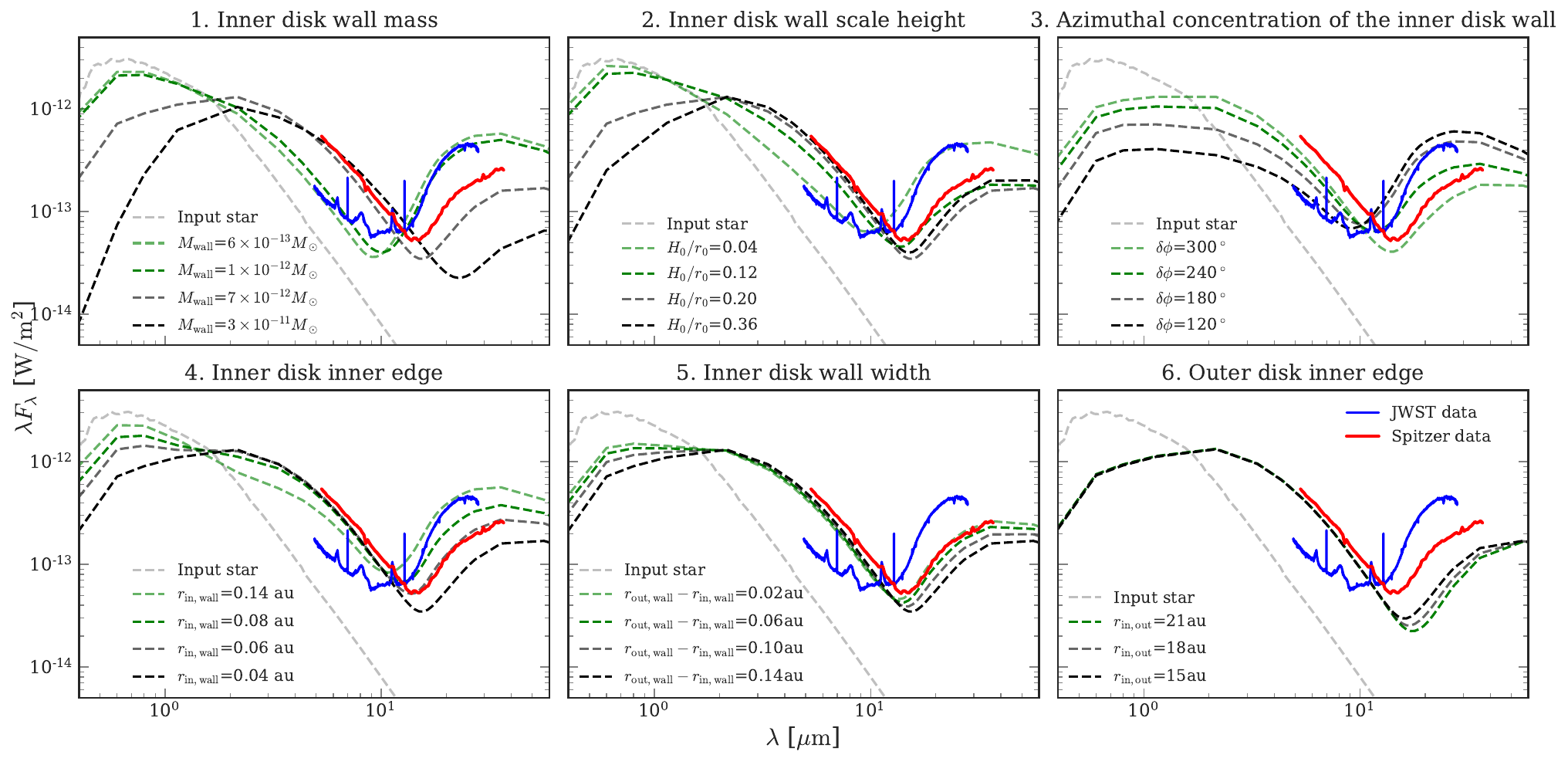}
    \caption{SED changes with each single parameter. When changing one parameter, other parameters are kept fixed as shown in Table~\ref{app:tab:Start}. In Panel~3 the optically thick part is between the star and the observer to illustrate how the maximum optical extinction vary with $\delta \phi$. Notice that the initial parameters set here are not exactly the same as our fitted model for {\it Spitzer}. 
    }
    \label{app:fig:single}
\end{figure*}

Our results are shown in Figure~\ref{app:fig:single}. The observed SEDs from {\it Spitzer} and JWST are also shown. The summary of how SED changes with each parameter is as follows:

\textbf{Panel 1. Inner disk wall dust mass.} For disks with a large gap separating the low-mass inner disk and the outer disk like T~Cha, changing the inner disk wall mass is the easiest way to reproduce the seesaw behavior at mid-IR bands. Reducing the dust mass in the inner disk wall can transform the optically thick inner disk wall into optically thin, resulting in less emission at shorter wavelengths ($<10\mu m$ which is from the inner disk) and more emission at longer wavelengths ($>10\mu m$ for more stellar radiation can reach the outer disk). This can lead to less optical extinction of the star from the disk as well. 
%The grain size distribution is degenerate with the inner disk mass for mid-IR SED modelling because mid-IR emission is mostly from the smaller grains. But the change in grain size distribution will not influence the stellar optical extinction. 
We also notice that the inner disk wall mass of $7\times 10^{-11} M_\odot$ (for the grain size distribution we selected) is at a critical stage between optically thick and thin, and slightly varying the mass can significantly influence the mid-IR spectrum. 

\textbf{Panel 2. Inner disk wall scale height.} Changing the inner disk wall scale height is another way of changing the height of the optically thick part of the inner disk. Instead of reducing the dust mass, scale height can change the vertical distribution of it. For disks similar to T~Cha, 
%where the inner disk wall is at a critical stage between optically thick and thin, 
only reducing the inner disk scale height to less than 0.1 (which coincides with the outer disk scale height in our model) can lead to significant seesaw in mid-IR bands (see the light blue line compared with dark blue and black lines in Fig~\ref{app:fig:single}). 

\textbf{Panel 3. Azimuthal concentration of the inner disk wall.} In our model, we find that changing $\delta \phi$ can also result in significant mid-IR seesaw behavior. Compared with changing scale height which influence the vertical distribution of the dust mass, $\delta \phi$ indicates the azimuthal distribution of it. Concentrating the dust mass azimuthally to a smaller region can lead to less emission from the inner disk while allowing more stellar radiation to reach the outer disk, similar to reducing the scale height which concentrates the dust mass vertically to a smaller region. Reducing $\delta \phi$ also increases the maximum optical extinction while reducing the time of the extinction phase in each period of the optical variability. 

The three parameters mentioned above can significantly influence the mid-IR SED and lead to the strong seesaw behavior. Among all other parameters, only the inner edge of the inner disk wall can lead to a slight seesaw behavior (see Panel.4 of Fig~\ref{app:fig:single}). Other parameters, including the wall width ($r_{\rm out, wall}-r_{\rm in, wall}$), surface density distribution exponent ($\alpha$) and flaring exponent ($\beta$), will hardly influence the shorter wavelength emission for disks like T~Cha (see Panel.5 of of Fig~\ref{app:fig:single} as an example). As a ``wall'', the width of it should be low and thus the radial mass distribution and structure can only have very little influence on the emission. 

Outer disk parameters, on the other hand, only influence the longer wavelength emission ($>10\mu m$, see Panel.6 of Fig~\ref{app:fig:single} as an example). The effect of varying other disk parameters has been well studied by \cite{Woitke16}. 
%In this way, these parameters will not lead to the mid-IR seesaw behavior, and are beyond the scope of this paper. 

%We find that changing the dust mass, scale height and the azimuthal distribution of dust mass ($\delta \phi$) can significantly influence the mid-IR SED, and reproducing the seesaw behavior. 

\begin{figure*}[htb!]
    \centering
	\includegraphics[width=0.49\textwidth]{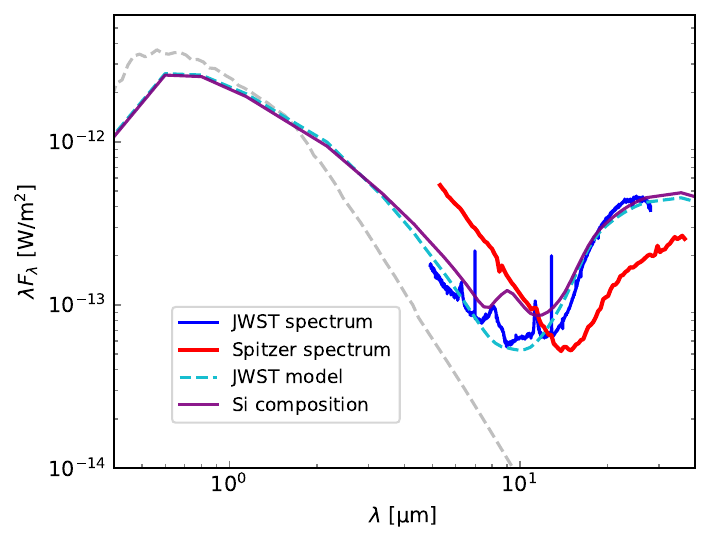}
    \caption{Model for inner disk without a wall (purple) compared with the model of a wall $+$ hidden mass. We can see that the model without a wall will have a strong 10~$\mu$m silicate feature, which does not match our observed data. 
    }
    \label{app:fig:Si}
\end{figure*}

In addition, we also present here the influence on the SED from adding small Silicate grains in the inner disk (see Fig~\ref{app:fig:Si}). In this model, we remove the inner disk wall, and leave the previously hidden mass with the same composition as the outer disk (70\% of silicate and 30\% of graphite) as the main emission source of the NIR wavelength. We vary other parameters to get a relatively good fit with our JWST spectrum. However, the strong 10~$\mu$m Si feature in this case does not match our data, meaning that the inner disk wall cannot be completely destroyed. 

\section{Photometry Table} 
Table~\ref{app:tab:ref} summarizing the references of all the photometric points we plot in Fig~\ref{fig:Overview} and \ref{fig:asym}.

\begin{deluxetable*}{cc}
%\tablenum{4}
\tablecaption{References of the photometric points. The photometric points used to model the SED (see Fig.~3) can also be downloaded here.}\label{app:tab:ref}
\tablewidth{1.2\textwidth}
\tablehead{
%\multicolumn{3}{c}{Fixed parameters}\\
%\hline
%Name & $\lambda$ ($\mu m$) & Flux ($W/m^2$)& Error ($W/m^2$) & Ref\\
Name & Ref\\
}
\startdata
Alcala 1989-1992 & \cite{Alcala93}\\
\multirow{2}{*}{AKARI stacked points} & FIS \cite{Akari10} \\
& IRC \cite{Akari10IRS} \\
Spitzer photometry & \cite{Spitzer03} \\
MSX 1996-1997 & \cite{MSX03} \\
2MASS 2000 & \cite{2MASS} \\
IRAS 1983 & \cite{IRAS94}\\
WISE 2010 & \cite{WISE12}\\
Walter 2018 & \cite{Walter18}\\
Herschel 2010 & \cite{Cieza11,Lommen07}\\
WISE 2022 & NEOWISE data release\\
ASAS-SN 2021-2023 & ASAS-SN data release\\
AAVSO & https://www.aavso.org\\
%Akari stacked points & 18.4 & 1.3e-13 & 8.5e-15 & a\\
%Akari stacked points & 8.6 & 2.1e-13 & 5.3e-15 & a\\
%Akari stacked points & 160 & 6.8e-14 & 4.1e-14 & a\\
%Akari stacked points & 140 & 1.0e-13 & 1.5e-14 & a\\
%Akari stacked points & 90 & 1.9e-13 & 5.7e-15 & a\\
%Akari stacked points & 64 & 2.6e-13 & 6.0e-15 & a\\
%Spitzer photometry & 7.9& 2.7e-13& 1.4e-14& b \\
%Spitzer photometry & 5.7& 5.9e-13&4.3e-14& b \\
%Spitzer photometry & 4.5& 8.9e-13&4.8e-14& b \\
%Spitzer photometry & 3.5& 1.2e-12&7.1e-14& b \\
\enddata

\end{deluxetable*}

\bibliography{Ref}
\bibliographystyle{aasjournal}



\end{document}